\documentclass[times,twocolumn,final]{elsarticle}

\usepackage{medima}
\usepackage{framed,multirow}

\usepackage{amssymb}
\usepackage{latexsym}

\usepackage{url}
\usepackage{xcolor}

\usepackage{hyperref}

\usepackage{booktabs}
\usepackage{multirow}

\usepackage{algorithm}
\usepackage{algpseudocode}
\usepackage[labelfont=bf,textfont=normalfont]{caption}

\journal{Under Review}

\begin{document}

\verso{Qiang Ma \textit{et~al.}}

\begin{frontmatter}

\title{Cardiac Mesh Flow: One-Step Generation of 3D+t Cardiac Four-Chamber Meshes via Flow Matching}


\author[1,2]{Qiang Ma\corref{cor1}}
\cortext[cor1]{Corresponding author.}
\ead{q.ma20@imperial.ac.uk}
\author[2,3]{Qingjie Meng}
\author[2,4]{Mengyun Qiao}
\author[1,5,6]{Paul M. Matthews}
\author[7]{Declan P. O'Regan}
\author[1,2,8]{Wenjia Bai}

\address[1]{Department of Brain Sciences, Imperial College London, London, UK}
\address[2]{Department of Computing, Imperial College London, London, UK}
\address[3]{School of Computer Science, University of Birmingham, Birmingham, UK}
\address[4]{Department of Mechanical Engineering, University College London, London, UK}
\address[5]{UK Dementia Research Institute, Imperial College London, London, UK}
\address[6]{Rosalind Franklin Institute, Harwell Science and Innovation Campus, Didcot, UK}
\address[7]{MRC Laboratory of Medical Sciences, Imperial College London, London, UK}
\address[8]{Data Science Institute, Imperial College London, London, UK}

\received{30 April 2026}

\begin{abstract}

Spatio-temporal (3D+t) generative modelling of cardiac shape and motion is crucial for understanding heart structure and function at population scale. Existing generative models for cardiac shape synthesis either adopt volumetric shape representations that lack anatomical correspondence across different time points and subjects, or rely on VAE-based frameworks that suffer from a trade-off between reconstruction fidelity and generative diversity. In this work, we propose Cardiac Mesh Flow, a novel generative flow model for 3D+t cardiac four-chamber mesh generation with anatomical correspondence, temporal coherence, and periodic consistency. Leveraging the flow matching technique, Cardiac Mesh Flow performs efficient one-step generation of multi-scale free-form deformation fields, which warp a template mesh to generate cardiac four-chamber meshes across a cardiac cycle. Furthermore, Cardiac Mesh Flow enables controllable generation conditioned on cardiac chamber volumes, allowing precise control of the synthetic heart. Experimental results demonstrate that Cardiac Mesh Flow achieves high fidelity and diversity on both unconditional and conditional generation, compared to state-of-the-art 3D+t cardiac mesh generation methods.

\end{abstract}

\begin{keyword}
\KWD  Cardiac imaging  \sep Anatomical shape modelling \sep Generative models \sep Flow matching
\end{keyword}

\end{frontmatter}


\section{Introduction}
Spatio-temporal (3D+t) cardiac shape modelling is a fundamental research question in cardiac imaging, as both the structural and motion patterns are closely associated with cardiac health and diseases. Generative models provide a flexible and compact framework for modelling the underlying distribution of cardiac shape and motion, facilitating the characterisation of heart variation in large populations \citep{xia2022msci,burns2024genetic,qiao2025meshheart}, the synthesis of realistic anatomical shapes \citep{sorensen2024stndf,ma2025cardiacflow,dou2025cardiosynth}, and the personalisation of cardiac digital twins \citep{corral2020digital,niederer2021digital,qian2025digital}.

Current generative models characterise cardiac geometry based on either volumetric shape representations \citep{qiao2023cheart,yang2024imheart,kong2024sdf4chd,sorensen2024stndf,ma2025cardiacflow}, such as segmentation maps and signed distance functions, or explicit surface meshes \citep{gaggion2025hybridvnet,qiao2025meshheart,dou2025cardiosynth}. In terms of volumetric shape representation, \cite{yang2024imheart} introduced a latent variable model, called ImHeart, to generate whole-heart occupancy fields by jointly learning an implicit template and deformation fields. \cite{kong2024sdf4chd} proposed SDF4CHD, a deep generative model to learn signed distance fields for synthesising cardiac anatomies with congenital heart diseases. Beyond modelling static cardiac structures, \cite{sorensen2024stndf} proposed a 3D+t neural distance field to learn both the shape and motion of the left atrium. However, volumetric shape representations, such as segmentations and signed distance fields, are unable to provide explicit geometric parameterisation with anatomical correspondence. As a result, additional post-processing steps are required to align cardiac structures across different subjects or different time frames of a cardiac cycle.

Instead of modelling the volumetric shapes of the heart, recent approaches \citep{gaggion2025hybridvnet,qiao2025meshheart,dou2025cardiosynth} learn to synthesise both cardiac shape and motion represented by surface meshes, which provide not only parameter-efficient shape representations, but also vertex-wise correspondence across different time frames and subjects, enabling time-resolved motion tracking as well as inter-subject analysis. \cite{gaggion2025hybridvnet} introduced a variational autoencoder (VAE) \citep{kingma2014vae} for 3D cardiac four-chamber mesh generation, named as HybridVNet, which consists of an encoder for multi-view CMR images and a decoder for mesh generation with multi-resolution graph convolutional networks (GCN) \citep{kipf2017gcn}. \cite{qiao2025meshheart} proposed MeshHeart, a VAE-based conditional generative model that incorporates a GCN encoder and a temporal Transformer architecture to generate 3D+t cardiac mesh sequences. \cite{dou2025cardiosynth} proposed the 4D CardioSynth method for dynamic cardiac mesh synthesis, which employs a VAE model with GCN layers and disentangles the latent space for shape reconstruction and motion tracking respectively.

Regardless of different shape representations of the heart, most existing approaches for cardiac shape generation mainly rely on VAE-based generative models \citep{kingma2014vae}, which often face a trade-off between reconstruction fidelity and latent space diversity and tend to over-smooth the shape and motion of the generated heart. Recently, flow matching approaches have received increasing attention for generative modelling \citep{lipman2023fm,liu2023rectified,esser2024stable3,lee2024flow}. Compared to the VAE, flow matching avoids the reconstruction–diversity trade-off and learns a continuous flow from noise to data, leading to high generation fidelity and diversity. Leveraging a latent flow matching technique, \cite{ma2025cardiacflow} proposed CardiacFlow for the synthesis of 3D+t cardiac four-chamber segmentation maps. An efficient one-step generation is achieved by jointly learning the initial values and velocity fields of the flow model.

\begin{figure*}[t]
\centering
\includegraphics[width=0.96\linewidth]{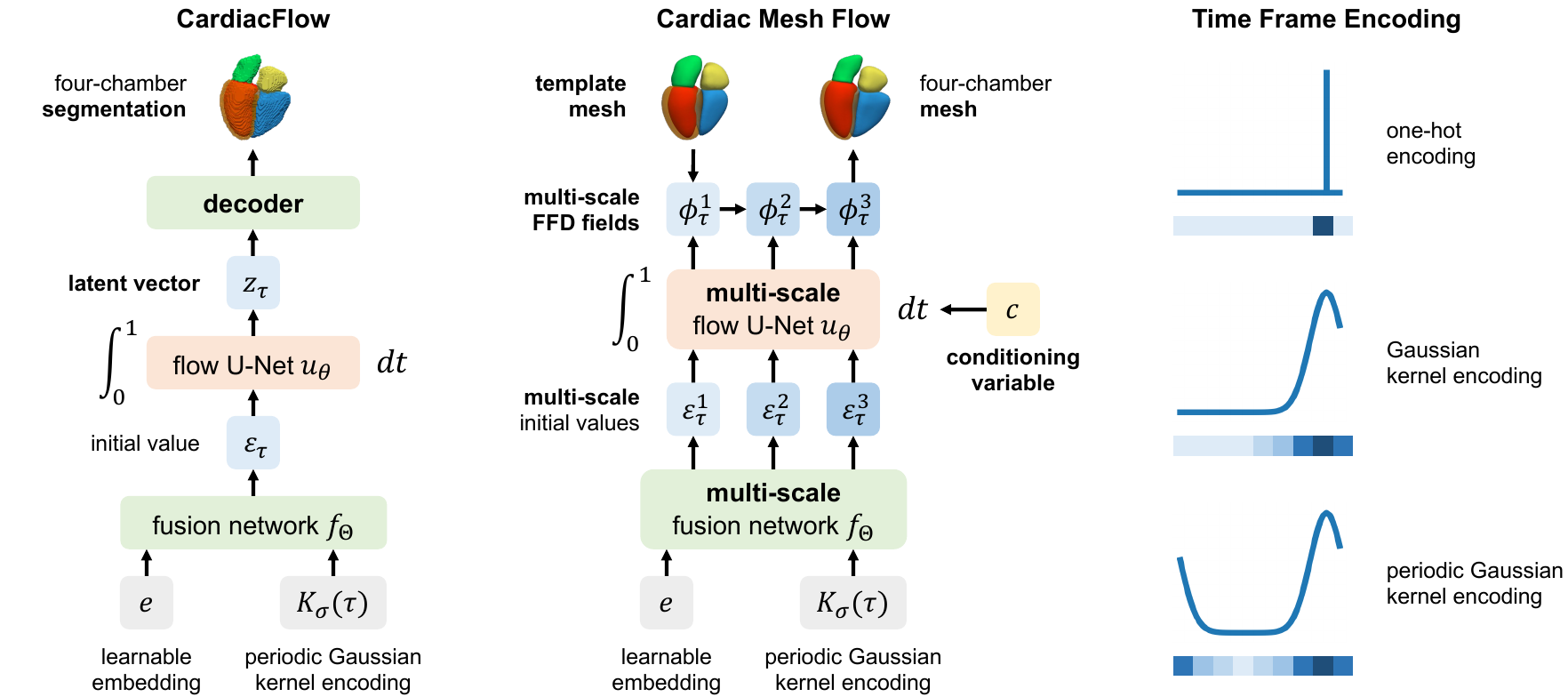}
\caption{Comparison between CardiacFlow~\citep{ma2025cardiacflow} and Cardiac Mesh Flow. Left: CardiacFlow uses a one-step generative flow to learn a latent distribution, which is then decoded into a four-chamber segmentation map. Middle: Cardiac Mesh Flow employs a fusion network to predict multi-scale initial values as the inputs of a flow U-Net to generate multi-scale FFD fields. The FFD fields warp a template mesh into synthetic cardiac four-chamber mesh. By incorporating temporal information of each time frame, Cardiac Mesh Flow can generate cardiac meshes across a cardiac cycle. Right: comparison across different temporal encoding strategies for a time frame. The temporal encoding is a $N$-dimensional vector, where $N$ denotes the total number of time frames in a cardiac cycle. }
\label{fig:cmflow}
\end{figure*}

As an extension to CardiacFlow presented at MICCAI 2025~\citep{ma2025cardiacflow}, here, we present \textit{Cardiac Mesh Flow}, a one-step flow matching approach for 3D+t cardiac four-chamber mesh generation with anatomical correspondence. As illustrated in Fig.~\ref{fig:cmflow}, Cardiac Mesh Flow introduces a multi-scale flow matching framework to generate multi-scale FFD fields that capture the shape and motion variation of the heart. The four-chamber cardiac mesh of each time frame is generated by warping a template mesh using the synthetic FFD fields. The temporal position of each time frame is encoded by \textit{periodic Gaussian kernel encoding}, which enforces temporal and periodic consistency on the generated 3D+t meshes. Furthermore, Cardiac Mesh Flow enables controllable generation of 3D+t cardiac meshes conditioned on the chamber volumes. The code of this work is publicly released at \url{https://github.com/m-qiang/CardiacMeshFlow}.

Compared to the MICCAI 2025 publication~\citep{ma2025cardiacflow}, Cardiac Mesh Flow substantially extends the methodology and experimental results in the following aspects:

\begin{itemize}
\item While CardiacFlow generates the latent vectors of 3D+t cardiac four-chamber segmentations, Cardiac Mesh Flow extends the method to generate 3D+t cardiac four-chamber meshes with vertex-wise anatomical correspondence.

\item Cardiac Mesh Flow leverages a multi-scale flow matching framework for efficient one-step generation of multi-scale FFD fields, which warp a template mesh to synthesise cardiac four-chamber meshes at each time frame.

\item Cardiac Mesh Flow introduces controllability to the generation process by incorporating conditioning variables, such as chamber volumes, for precise control of the generated cardiac four-chamber meshes.
\end{itemize}

\section{Cardiac Mesh Flow}
\textit{Cardiac Mesh Flow} is a one-step generative flow that extends the CardiacFlow approach \citep{ma2025cardiacflow} to generate multi-scale FFD fields for 3D+t cardiac four-chamber mesh synthesis. As illustrated in Fig.~\ref{fig:cmflow}, Cardiac Mesh Flow fuses a learnable embedding $e$ and a \textit{periodic Gaussian kernel encoding} of the time frame $\tau$ to predict multi-scale learnable initial values for a flow model. Based on the initial values, a flow U-Net $u_{\theta}$ is trained to generate multi-scale FFD fields $\phi^l$ for $l=1,...,L$. The four-chamber mesh at time frame $\tau$ is generated by apply the multi-scale FFD fields to warp a template mesh. A dynamic cardiac mesh sequence is generated by iterating $\tau=1,...,N$, where $N$ denotes the number of time frames in a cardiac cycle.

In this section, we first briefly introduce the HeartFFDNet \citep{ma2026heartssm}, which learns multi-scale FFD fields for cardiac mesh modelling. Then, we review the necessary background for flow matching \citep{lipman2023fm,liu2023rectified}, describe the multi-scale FFD generation framework and the one-step generative flow model \citep{ma2025cardiacflow}, and finally provide the algorithms for unconditional and conditional 3D+t four-chamber mesh generation using Cardiac Mesh Flow.

\subsection{Learning FFD for mesh representation}\label{sec:recon}
\begin{figure*}[!t]
\centering
\includegraphics[width=0.9\linewidth]{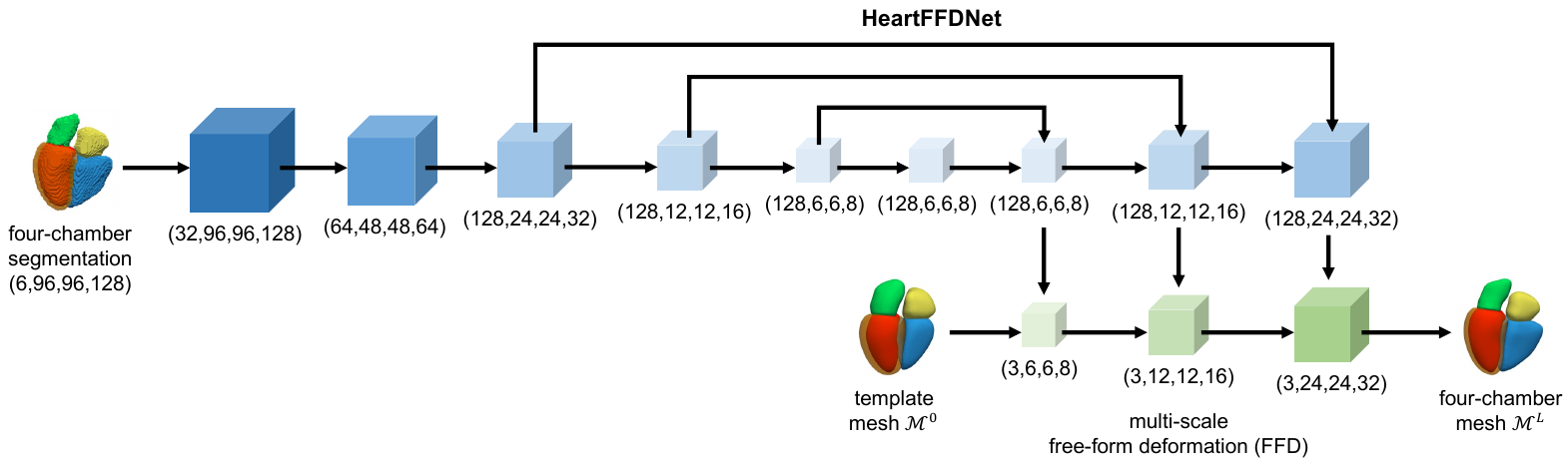}
\caption{The architecture of HeartFFDNet. HeartFFDNet learns to predict multi-scale free-form deformation (FFD) fields from an input 3D cardiac four-chamber segmentation map. The corresponding mesh is reconstructed by warping a template mesh according to the predicted FFD fields. The multi-scale FFD fields predicted by HeartFFDNet serve as the training data for Cardiac Mesh Flow.}
\label{fig:heartffdnet}
\end{figure*}

We utilise HeartFFDNet \citep{ma2026heartssm} to learn multi-scale FFD fields for the generative modelling of cardiac four-chamber meshes via Cardiac Mesh Flow. Fig.~\ref{fig:heartffdnet} illustrates the architecture of the HeartFFDNet. The input of HeartFFDNet is a 3D segmentation map of the heart, including the left ventricle (LV), LV myocardium (LVM), right ventricle (RV), left atrium (LA), and right atrium (RA). The outputs are multi-scale FFD fields $\phi^l$ for scale $l=1,...,L$ with $L=3$. A four-chamber mesh $\mathcal{M}^L=(\mathcal{V}^L,\mathcal{E}^L,\mathcal{F}^L)$ is reconstructed by applying the multi-scale FFD fields $\phi^l$ to warp a template mesh $\mathcal{M}^0=(\mathcal{V}^0,\mathcal{E}^0,\mathcal{F}^0)$ via B-spline interpolation, \emph{i.e.}, $\mathcal{M}^l=\mbox{B-spline}(\mathcal{M}^{l-1},\phi^l)$ for $l=1,...,L$. The template four-chamber mesh consists of five structures: LV endocardium (LV-endo), LV epicardium (LV-epi), RV, LA, and RA. 

HeartFFDNet is trained by minimising a structure-wise combination of a Chamfer distance $\mathcal{L}_{\mathrm{cd}}$, a mesh edge loss $\mathcal{L}_{\mathrm{edge}}$, and a mesh curvature loss $\mathcal{L}_{\mathrm{curv}}$ \citep{ma2026heartssm}. The pseudo ground-truth point cloud $\mathcal{V}^*$ is extracted from the input segmentation using the marching cubes algorithm \citep{lorensen1998marching}. Given the vertices $\mathcal{V}^L$ of the predicted mesh $\mathcal{M}^L$, the Chamfer distance between $\mathcal{V}^L$ and $\mathcal{V}^*$ is defined as:
\begin{equation}\label{eq:ffdnet-loss-chamfer}
\mathcal{L}_{\mathrm{cd}}=\sum_{v^L\in\mathcal{V}^L}\min_{v^*\in\mathcal{V}^*}\|v^L-v^*\|^2+\sum_{v^*\in\mathcal{V}^*}\min_{v^L\in\mathcal{V}^L}\|v^*-v^L\|^2.
\end{equation}
The mesh edge loss $\mathcal{L}_{\mathrm{edge}}$ is defined as the standard deviation of the edge length. The mesh curvature loss measures the discrepancy of the mean curvatures between the predicted mesh $\mathcal{M}^L$ and the template mesh $\mathcal{M}^0$:
\begin{equation}\label{eq:ffdnet-loss-curv}
\mathcal{L}_{\mathrm{curv}}=1-r(H^L,H^0),
\end{equation}
where $r(\cdot,\cdot)$ is the Pearson correlation coefficient, $H^L$ and $H^0$ are the mean curvatures of all vertices of the meshes $\mathcal{M}^L$ and $\mathcal{M}^0$ respectively. Such a curvature loss encourages the vertex correspondence of the four-chamber meshes across different subjects and time frames. The final loss function is defined as
\begin{equation}\label{eq:ffdnet-loss-all}
\mathcal{L}=\mathcal{L}_{\mathrm{cd}}+w_{\mathrm{edge}}\mathcal{L}_{\mathrm{edge}}+w_{\mathrm{curv}}\mathcal{L}_{\mathrm{curv}},
\end{equation}
where $w_{\mathrm{edge}}$ and $w_{\mathrm{curv}}$ are the weights of the mesh regularisation terms. The loss function is computed individually for each of the five cardiac structures and then summed together. Once trained, HeartFFDNet allows the cardiac mesh at each time frame $\tau$ to be represented by multi-scale FFD fields, enabling the generative modelling of cardiac four-chamber meshes.

\subsection{Flow matching}\label{sec:fm}
The \textit{flow matching} approach aims to learn a flow trajectory $z_t$ for $t\in[0,1]$ from a prior distribution $p_{\mathrm{prior}}(\varepsilon)$ to a data distribution $q_{\mathrm{data}}(x)$. The trajectory $z_t$ is modelled by a flow ordinary differential equation (ODE):
\begin{equation}\label{eq:fm-ode}
\frac{\mathrm{d}z_t}{\mathrm{d}t}=u_{\theta}(z_t;t),~z_0\sim p_{\mathrm{prior}},~t\in[0,1],
\end{equation}
where $u_{\theta}$ is a time-varying velocity field parameterised by a deep neural network model, and $z_0=\varepsilon\sim p_{\mathrm{prior}}$ is an initial value. To optimise the neural network model $u_{\theta}$, a standard approach is to model the trajectory $z_t$ as the linear interpolation between prior $\varepsilon\sim p_{\mathrm{prior}}$ and data $x\sim q_{\mathrm{data}}$, \emph{i.e.},
\begin{equation}\label{eq:fm-linear}
z_t=(1-t)\varepsilon+tx.
\end{equation}
In this case, the velocity field $u_{\theta}$ is learnt such that the flow path $z_t$ follows a straight line with velocity $u=x-\varepsilon$. Consequently, the flow matching loss is defined as:
\begin{equation}\label{eq:fm-loss}
\mathcal{L}(\theta)=\mathbb{E}_{t,\varepsilon,x}\left[\|u_{\theta}(z_t;t)-(x-\varepsilon)\|^2\right].
\end{equation}
During training, we randomly sample the prior $\varepsilon\sim p_{\mathrm{prior}}$, the data $x\sim q_{\mathrm{data}}$, and sample the time $t$ from a uniform distribution $\mathcal{U}(0,1)$. The input $z_t$ of the flow network $u_{\theta}$ is computed according to Eq.~(\ref{eq:fm-linear}). Such a flow model is also called a \textit{rectified flow} \citep{liu2023rectified,esser2024stable3}.

After training, we can generate new data samples by solving the flow ODE (\ref{eq:fm-ode}) using integration schemes such as the forward Euler method:
\begin{equation}\label{eq:fm-euler}
z_{k+1}=z_k+hu_{\theta}(z_k;kh),
\end{equation}
where $k=0,...,K-1$ is the integration step and $h=1/K$ is the step size. Note that if $u_{\theta}$ is learnt such that the flow path $z_t$ exactly follows a straight line, the flow ODE (\ref{eq:fm-ode}) can be solved by only one forward Euler step:
\begin{equation}\label{eq:fm-euler}
z_{1}=z_0+u_{\theta}(z_0;0).
\end{equation}

\begin{figure*}[!t]
\centering
\includegraphics[width=0.9\linewidth]{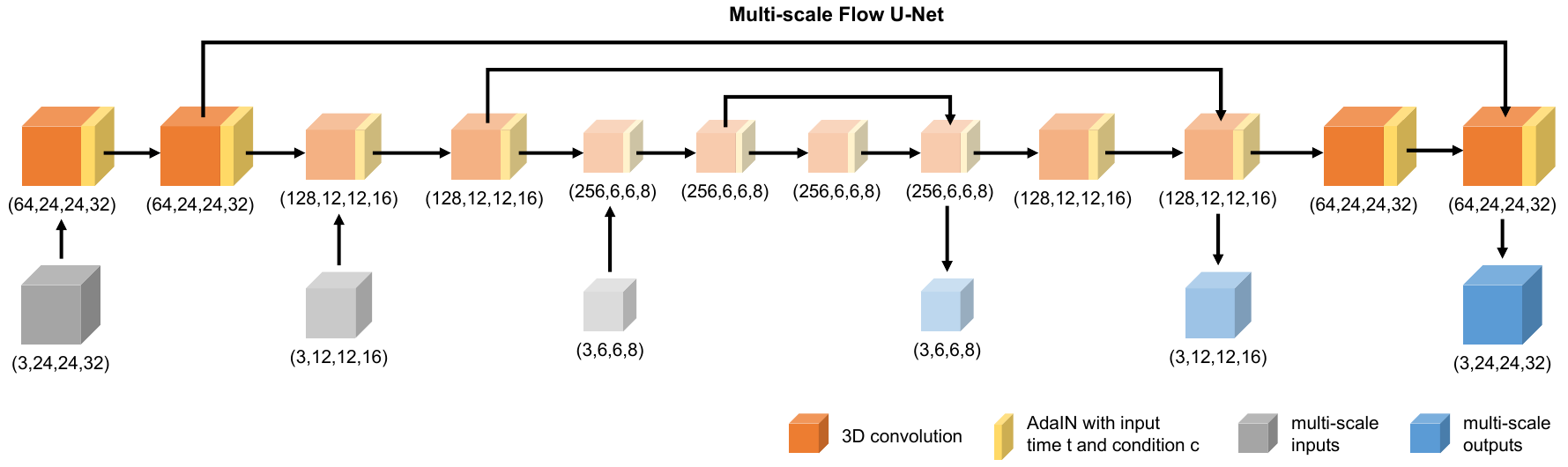}
\caption{The architecture of multi-scale flow U-Net. The flow U-Net consists of 3D convolutional blocks following adaptive instance normalisation (AdaIN) layers, which capture the features of the integration time $t$ and conditioning variable $c$. The size of the multi-scale inputs and outputs correspond to the multi-scale FFD fields required for cardiac four-chamber mesh reconstruction.}
\label{fig:flowunet}
\end{figure*}

\subsection{Multi-scale FFD generation}
Cardiac Mesh Flow generates multi-scale FFD fields to warp a template mesh into a cardiac four-chamber mesh at each time frame $\tau$. For any multi-scale initial values $\varepsilon^l_{\tau}\sim p^l_{\mathrm{prior}}$ and multi-scale FFD fields $\phi^l_{\tau}\sim q^l_{\mathrm{data}}$ at frame $\tau$ for scale $l=1,...,L$, we introduce a multi-scale flow matching framework modelled by the following ODE system:
\begin{equation}\label{eq:cmflow-ode}
\frac{\mathrm{d}z^l_{t,\tau}}{\mathrm{d}t}=u^l_{\theta}\left(z^1_{t,\tau},...,z^L_{t,\tau};t\right),~z^l_{0,\tau}\sim p^l_{\mathrm{prior}},~t\in[0,1],
\end{equation}
where $u_{\theta}$ is parameterised by a \textit{multi-scale flow U-Net}, and $u^l_{\theta}$ is the output of $u_{\theta}$ at the $l$-th scale. As illustrated in Fig.~\ref{fig:flowunet}, the flow U-Net $u_{\theta}$ incorporates multi-scale inputs and outputs. Each convolutional layer is followed by a \textit{adaptive instance normalisation} (AdaIN) layer \citep{karras2019style} to encode the time $t$ of the flow and other conditioning variables.

At each scale $l=1,...,L$, we model the flow trajectory $z^l_{t,\tau}$ as the linear interpolation between the initial value $\varepsilon^l_{\tau}$ and the FFD field $\phi^l_{\tau}$:
\begin{equation}\label{eq:cmflow-linear}
z^l_{t,\tau}=(1-t)\varepsilon^l_{\tau}+t\phi^l_{\tau}.
\end{equation}
Then, a multi-scale flow matching loss can be defined by
\begin{equation}\label{eq:cmflow-loss}
\mathcal{L}(\theta)=\mathbb{E}_{t,\tau,\varepsilon,\phi}\left[\sum_{l=1}^L\left\|u^l_{\theta}\left(z^1_{t,\tau},...,z^L_{t,\tau};t\right)-(\phi^l_{\tau}-\varepsilon^l_{\tau})\right\|^2\right],
\end{equation}
where the ground truth multi-scale FFD fields $\phi^l_{\tau}$ are produced by the HeartFFDNet. After training, we generate the multi-scale FFD fields $\phi^l_{\tau}=z^l_{1,\tau}$ for $l=1,...,L$ by integrating the flow ODE system~(\ref{eq:cmflow-ode}). We apply the B-spline interpolation to the multi-scale FFD fields $\phi^l_{\tau}$, warping a template mesh $\mathcal{M}^0_{\tau}$ iteratively into the synthetic cardiac mesh $\mathcal{M}^L_{\tau}$ at time frame $\tau$.

\subsection{One-step generative flow}
Cardiac Mesh Flow is developed for one-step generation of the cardiac mesh at each time frame $\tau$. As discussed in Section \ref{sec:fm}, the crucial point for one-step generation is that the flow trajectory is learnt to be an ideal straight line, so that one step of forward Euler integration can accurately solve the flow ODE \citep{lipman2023fm,liu2023rectified}. Similarly to CardiacFlow \citep{ma2025cardiacflow}, one-step generation is achieved via straightening the flow path $z_t$ by jointly optimising both the initial value $\varepsilon$ and the flow model $u_{\theta}$. Specifically, as shown in Fig.~\ref{fig:cmflow}, we assign a learnable embedding $e$ to each training sample. A fusion network is learnt to predict multi-scale learnable initial values by fusing the learnable embedding $e$ and a temporal encoding of the time frame $\tau$. A \textit{periodic Gaussian kernel encoding} $\mathcal{K}_{\sigma}(\tau)$ is introduced to ensure the temporal and periodic consistency of the generated meshes.

\begin{figure}[!t]
\centering
\includegraphics[width=1.0\linewidth]{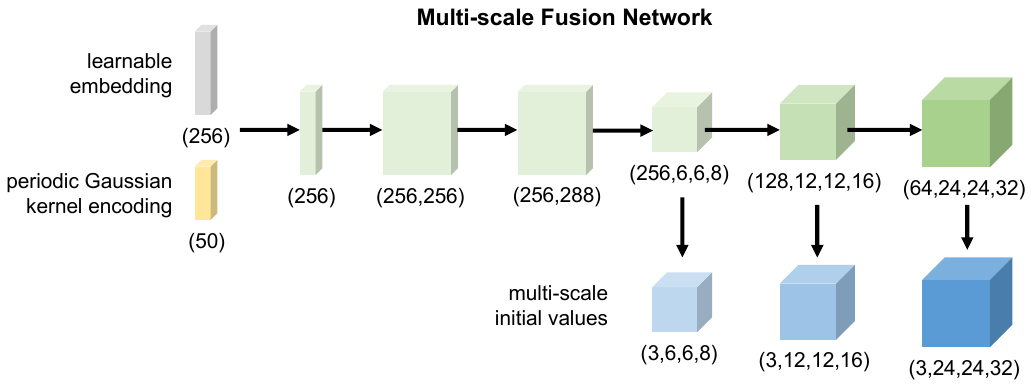}
\caption{The architecture of multi-scale fusion network. A learnable embedding and a periodic Gaussian kernel encoding of time frame are fused to predict multi-scale initial values for the flow ODE system.}
\label{fig:fusion}
\end{figure}

\paragraph{Multi-scale learnable initial values} 
The standard flow matching method utilises a normal distribution $\mathcal{N}(0,I)$ as the prior $p_{\mathrm{prior}}$ and samples random coupling of noise and data during training \citep{lipman2023fm,liu2023rectified}. To straighten the flow path for one-step generation, Cardiac Mesh Flow assigns a low-dimensional learnable embedding $e_{\phi}$ for each training sample $\phi\sim q_{\mathrm{data}}$, which is an FFD field. The embedding $e_{\phi}$ is independent of the scale $l$ and the time frame $\tau$. Subsequently, we propose a \textit{multi-scale fusion network} $f_{\Theta}$ to predict multi-scale frame-conditioned learnable initial values $\varepsilon^l_{\tau}$ from the learnable embedding and a temporal encoding $\mathcal{K}_{\sigma}(\tau)$ of the time frame $\tau$. In particular, we have
\begin{equation}\label{eq:fusion}
\left[\varepsilon^l_{\tau},...,\varepsilon^L_{\tau}\right]=f_{\Theta}\left(e_\phi, \mathcal{K}_{\sigma}(\tau)\right)
\end{equation}
The architecture of the multi-scale fusion network is presented in Fig.~\ref{fig:fusion}. For each time frame $\tau$, the fusion network provides multi-scale learnable initial values $z^l_{0,\tau}=\varepsilon^l_{\tau}$ for the flow ODE system (\ref{eq:cmflow-ode}). As a result, Cardiac Mesh Flow learns a straight flow trajectory $z^l_{t,\tau}$ for one-step generation by jointly optimising the learnable initial values $\varepsilon^l_{\tau}$ and the flow model $u_{\theta}$. For inference, we randomly sample the embedding $\hat{e}$ from an empirical Gaussian distribution $\mathcal{N}(\mu_e, \Sigma_e)$, which is derived from the learnable embeddings $\{e_{\phi}\}$ of all training samples.

\paragraph{Periodic Gaussian kernel encoding} Since cardiac motion is a dynamic process, Cardiac Mesh Flow generates cardiac mesh at each time frame $\tau$ conditioned on its temporal encoding. To enforce temporal and periodic consistency of the generated mesh sequence, we introduce a \textit{periodic Gaussian kernel} $\mathcal{K}_{\sigma}:\mathbb{R}\rightarrow\mathbb{R}^N$ to encode the time frame $\tau$ according to the periodicity of heartbeat \citep{ma2025cardiacflow}, \emph{i.e.}, the heart should have a consistent shape at the start ($\tau=1$) and end ($\tau=N$) of a cardiac cycle. For any time frame $\tau=1,...,N$, the encoding $\mathcal{K}_{\sigma}(\tau)$ is a $N$-dimensional vector with the $n$-th element defined by a Gaussian kernel:
\begin{equation}\label{eq:pgk-gauss}
\left[\mathcal{K}_{\sigma}(\tau)\right]_n:=\frac{1}{\sqrt{2\pi}\sigma}\exp\left(-\frac{d(n,\tau)^2}{2\sigma^2}\right),
\end{equation}
where $\sigma$ is the standard deviation, and the distance function $d(\cdot,\cdot)$ is formulated as:
\begin{equation}\label{eq:pgk-gauss}
d(n,\tau):=\min\left\{|n-\tau|, N-|n-\tau|\right\}.
\end{equation}
Such a distance metric is symmetric $d(n,\tau)=d(\tau,n)$ and periodic $d(n+N,\tau)=d(n,\tau)$ with a period of $N$. As shown in Fig.~\ref{fig:cmflow}-Right, the elements of the encoding $\mathcal{K}_{\sigma}(\tau)$ follow a truncated and periodic Gaussian distribution, while $[\mathcal{K}_{\sigma}(\tau)]_n$ achieves maximum when $n=\tau$. Compared to the one-hot encoding and standard Gaussian kernel encoding in Fig.~\ref{fig:cmflow}-Right, the proposed periodic Gaussian kernel method encodes neighbouring information of each frame $\tau$ continuously and periodically, thereby effectively improving the temporal and periodic consistency of the generated mesh sequence without explicit temporal regularisation.

\paragraph{Beta sampling} The classic flow matching approach samples time points $t$ from a uniform distribution \citep{lipman2023fm,liu2023rectified} or a logit-normal distribution \citep{esser2024stable3,lee2024flow} during training. To provide more samples at $t=0$ for one-step generation, we adopt \textit{Beta sampling} strategy \citep{lee2025beta} to sample more time points near $t=0$ from a Beta distribution $\beta(a,b)$ with $a=0.1$ and $b=2.0$, of which the density function decreases monotonically in $[0,1]$.

\algrenewcommand\algorithmicindent{1em}%
\begin{algorithm}[b]
\caption{Training (unconditional)}\label{alg:train-uncond}
\begin{algorithmic}[1]
\State \textbf{input:} data distribution $q_{\mathrm{data}}$, learnable embeddings $e_{\phi}$
\State \textbf{repeat}
\State \hspace{\algorithmicindent}
$\tau\sim\mathcal{U}(\{1,...,N\})$
\State \hspace{\algorithmicindent}
$\phi^1_{\tau},...,\phi^L_{\tau}\sim q_{\mathrm{data}}$
\State \hspace{\algorithmicindent}
$\varepsilon^1_{\tau},...,\varepsilon^L_{\tau}=f_{\Theta}\left(e_\phi, \mathcal{K}_{\sigma}(\tau)\right)$
\State \hspace{\algorithmicindent}
$t\sim\beta(a,b)$
\State \hspace{\algorithmicindent}
$z^l_{t,\tau}=(1-t)\varepsilon^l_{\tau}+t\phi^l_{\tau},~~l=1,...,L$
\State \hspace{\algorithmicindent}
$\mathcal{L}(\theta,\Theta,e_{\phi})=\sum_{l=1}^L\left\|u^l_{\theta}\left(z^1_{t,\tau},...,z^L_{t,\tau};t\right)-(\phi^l_{\tau}-\varepsilon^l_{\tau})\right\|^2$
\State \hspace{\algorithmicindent}
update $\theta$, $\Theta$, $e_{\phi}$
\end{algorithmic}
\end{algorithm}

\begin{algorithm}[b]
\caption{Generation (unconditional)} \label{alg:generate-uncond}
\begin{algorithmic}[1]
\State \textbf{input:} template mesh $\mathcal{M}^0$
\State $\hat{e}\sim\mathcal{N}(\mu_e,\Sigma_e)$
\For{$\tau=1,...,N$}
\State $\varepsilon^1_{\tau},...,\varepsilon^L_{\tau}=f_{\Theta}\left(\hat{e}, \mathcal{K}_{\sigma}(\tau)\right)$
\State $\phi^l_{\tau}=\varepsilon^l_{\tau}+u^l_{\theta}\left(\varepsilon^1_{\tau},...,\varepsilon^L_{\tau};0\right),~~l=1,...,L$
\State
$\mathcal{M}^0_{\tau} = \mathcal{M}^0$
\State
$\mathcal{M}^l_{\tau}=\mbox{B-spline}(\mathcal{M}^{l-1}_{\tau},\phi^l_{\tau}),~~l=1,...,L$
\EndFor
\State \textbf{return} four-chamber meshes $\mathcal{M}^L_{\tau},~~\tau=1,...,N$
\end{algorithmic}
\end{algorithm}

\begin{table*}[!t]
\centering
\setlength{\tabcolsep}{8pt}
\caption{Comparison results of 3D+t cardiac four-chamber mesh generation. Cardiac Mesh Flow is compared to state-of-the-art VAE-based cardiac mesh generation models and vanilla flow matching approaches. The unconditional generation ability is assessed by the volume FID score. The conditional generation performance is evaluated by the RMSD error between the input conditioning variables and four-chamber phenotypes of generated cardiac meshes. The runtime and GPU memory used for the inference are reported. The best results are highlighted in bold.}
\begin{tabular}{l|cccc}
\toprule
Method & vFID $\downarrow$ & $\mathrm{RMSD}_{\mathrm{cond}}$ $\downarrow$ &  Runtime & GPU memory \\
\midrule
HybridVNet \citep{gaggion2025hybridvnet} & $5.239$ & $6.303\pm3.444$ & $1.013$ s & $5.35$ GB\\
MeshHeart \citep{qiao2025meshheart}  & $5.035$ & $4.952\pm2.792$ & $\mathbf{0.003}$ s & $1.54$ GB \\
4D CardioSynth \citep{dou2025cardiosynth}  & $4.431$ & $6.445 \pm 2.635$ & $0.705$ s & $2.48$ GB \\
Flow Matching (1-step) & $2.827$ & $4.249\pm1.899$ & $0.347$ s & $\mathbf{1.23}$ GB\\
Flow Matching (5-step) & $2.524$ & $4.648\pm1.958$ & $1.028$ s & $\mathbf{1.23}$ GB\\
Cardiac Mesh Flow (Ours) & $\mathbf{0.439}$ & $\mathbf{2.816}\pm\mathbf{1.877}$ & $0.440$ s & $1.27$ GB\\
\bottomrule
\end{tabular}
\label{tab:compare}
\end{table*}

\subsection{3D+t cardiac mesh generation}
\paragraph{Unconditional generation} Incorporating the one-step generative flow into the multi-scale FFD generation framework, Cardiac Mesh Flow is able to synthesise realistic 3D+t cardiac four-chamber mesh sequence in one step of inference for each time frame.
The training and sampling procedures of Cardiac Mesh Flow are provided in Algorithms \ref{alg:train-uncond} and \ref{alg:generate-uncond}. Since Cardiac Mesh Flow generates cardiac meshes at different time frames independently, given sufficient GPU memory, we can generate a complete 3D+t mesh sequence in one step by setting the batch size to the total number of time frames $N$.

\paragraph{Controllable conditional generation} While previous studies have considered heart generation conditioned on demographic information \citep{sorensen2024stndf,qiao2025meshheart}, we demonstrate the conditional generation ability of Cardiac Mesh Flow using conditioning variables such as four-chamber volumes. For controllable conditional generation, we involve the conditioning vector $c$ into the multi-scale flow U-Net $u_{\theta}(\cdot;t,c)$, where the conditioning variable $c$ is captured by the AdaIN layers. We substitute the learnable embedding $e_{\phi}$ with random Gaussian noise $e\sim\mathcal{N}(0,I)$ for conditional generation, as the learnable embeddings have already contained rich information of all training cardiac shapes, which could limit the flow model $u_{\theta}$ to capture the conditioning information. The conditional generation procedure is provided in Algorithms \ref{alg:train-cond} and \ref{alg:generate-cond}.

\begin{algorithm}[h]
\caption{Training (conditional)}\label{alg:train-cond}
\begin{algorithmic}[1]
\State \textbf{input:} data distribution $q_{\mathrm{data}}$, conditioning variable $c$
\State \textbf{repeat}
\State \hspace{\algorithmicindent}
$\tau\sim\mathcal{U}(\{1,...,N\})$
\State \hspace{\algorithmicindent}
$\phi^1_{\tau},...,\phi^L_{\tau}\sim q_{\mathrm{data}}$
\State \hspace{\algorithmicindent}
$e\sim\mathcal{N}(0,I)$
\State \hspace{\algorithmicindent}
$\varepsilon^1_{\tau},...,\varepsilon^L_{\tau}=f_{\Theta}\left(e, \mathcal{K}_{\sigma}(\tau)\right)$
\State \hspace{\algorithmicindent}
$t\sim\beta(a,b)$
\State \hspace{\algorithmicindent}
$z^l_{t,\tau}=(1-t)\varepsilon^l_{\tau}+t\phi^l_{\tau},~~l=1,...,L$
\State \hspace{\algorithmicindent}
$\mathcal{L}(\theta,\Theta)=\sum_{l=1}^L\left\|u^l_{\theta}\left(z^1_{t,\tau},...,z^L_{t,\tau};t,c\right)-(\phi^l_{\tau}-\varepsilon^l_{\tau})\right\|^2$
\State \hspace{\algorithmicindent}
update $\theta$, $\Theta$
\end{algorithmic}
\end{algorithm}

\begin{algorithm}[h]
\caption{Generation (conditional)} \label{alg:generate-cond}
\begin{algorithmic}[1]
\State \textbf{input:} template mesh $\mathcal{M}^0$, conditioning variable $c$
\State $e\sim\mathcal{N}(0,I)$
\For{$\tau=1,...,N$}
\State $\varepsilon^1_{\tau},...,\varepsilon^L_{\tau}=f_{\Theta}\left(e, \mathcal{K}_{\sigma}(\tau)\right)$
\State $\phi^l_{\tau}=\varepsilon^l_{\tau}+u^l_{\theta}\left(\varepsilon^1_{\tau},...,\varepsilon^L_{\tau};0,c\right),~~l=1,...,L$
\State
$\mathcal{M}^0_{\tau} = \mathcal{M}^0$
\State
$\mathcal{M}^l_{\tau}=\mbox{B-spline}(\mathcal{M}^{l-1}_{\tau},\phi^l_{\tau}),~~l=1,...,L$
\EndFor
\State \textbf{return} four-chamber meshes $\mathcal{M}^L_{\tau},~~\tau=1,...,N$
\end{algorithmic}
\end{algorithm}

\section{Experiments}
In this section, we demonstrate the generation ability of Cardiac Mesh Flow on both unconditional and conditional 3D+t cardiac four-chamber mesh synthesis.

\subsection{Experimental settings}

\begin{figure*}[h]
\centering
\includegraphics[width=1.0\linewidth]{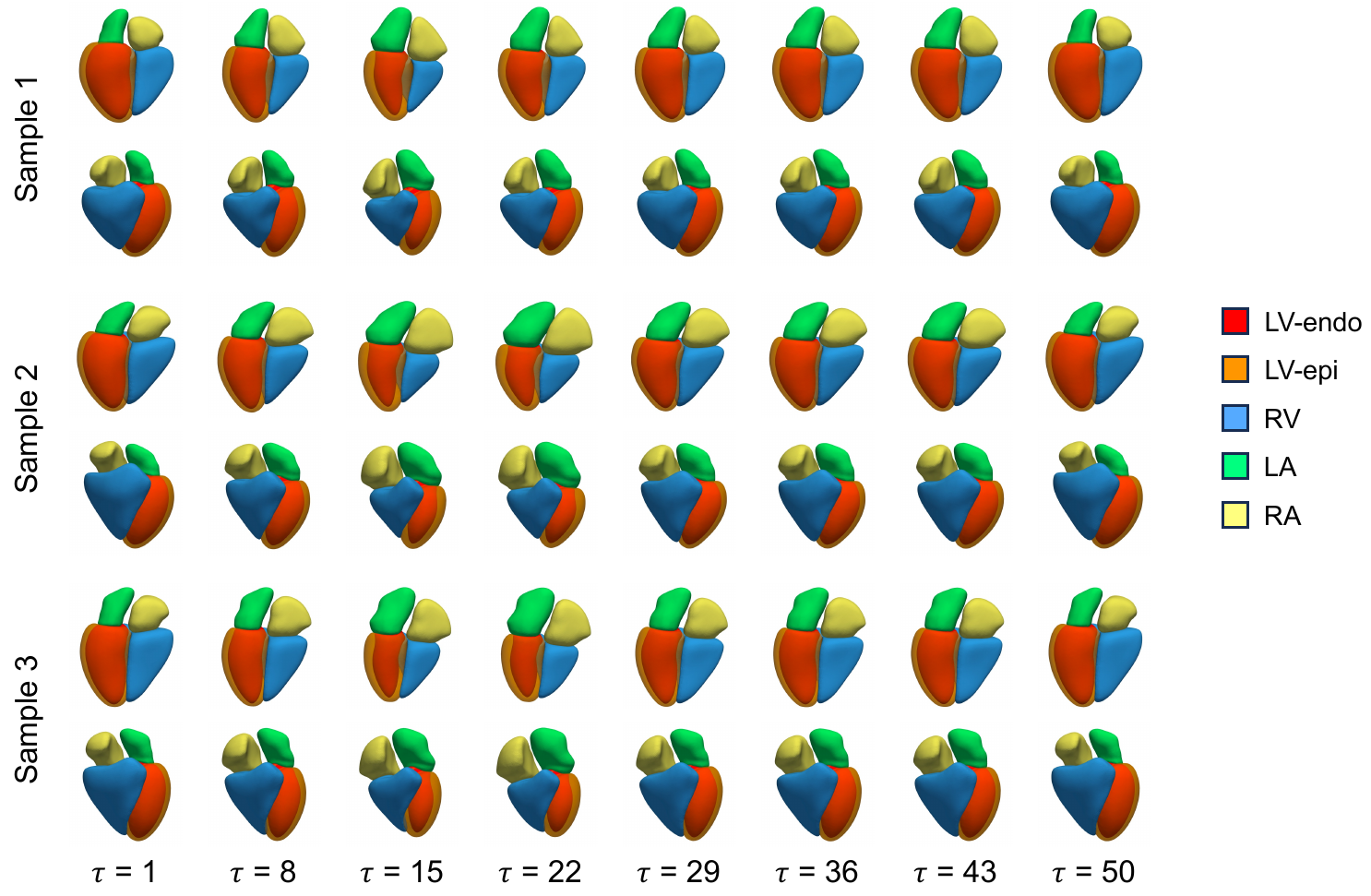}
\caption{Examples of 3D+t cardiac four-chamber meshes generated by Cardiac Mesh Flow.}
\label{fig:uncond}
\end{figure*}

\paragraph{Dataset} HeartFFDNet and Cardiac Mesh Flow is developed using a set of 1,200 multi-view CMR image sequences from the UK Biobank dataset \citep{petersen2016ukbb,bai2020population}, each sequence consisting of $N=50$ time frames. The four-chamber CMR segmentations are created by a public model with manual quality control \citep{bai2018automated}. The 3D+t dense segmentations are completed by a label completion U-Net \citep{xu2023complete,xu2024complete,muffoletto2024complete,ma2025cardiacflow}, which is trained following the settings in \cite{ma2026heartssm}. The dataset is split into 600/100/500 for training/validation/test. Following the procedure in Section \ref{sec:recon}, we employ HeartFFDNet to prepare the multi-scale FFD fields as the training data for Cardiac Mesh Flow.

\paragraph{Baselines} We compare Cardiac Mesh Flow with state-of-the-art (SOTA) explicit 3D+t cardiac mesh generation methods, including HybridVNet \citep{gaggion2025hybridvnet}, MeshHeart \citep{qiao2025meshheart}, and 4D CardioSynth \citep{dou2025cardiosynth}. For HybridVNet, we replace the input multi-view CMR images with 3D four-chamber segmentation maps to validate its generation ability. In addition, we compare to vanilla flow matching approach \citep{lipman2023fm,liu2023rectified} with different inference steps, using the same multi-scale neural network models as the Cardiac Mesh Flow.

\paragraph{Hyperparameters} For HeartFFDNet, we set $w_{\mathrm{edge}}=0.5$ and $w_{\mathrm{curv}}=1.0$ for the loss function (\ref{eq:ffdnet-loss-all}). For Cardiac Mesh Flow, we set $\sigma=1$ for periodic Gaussian Kernel encoding $\mathcal{K}_{\sigma}$. The number of resolution scales is set to $L=3$. All experiments are conducted on an Nvidia RTX 3090 GPU with 24GB memory.

\subsection{Unconditional generation}
\paragraph{Evaluation metrics} For unconditional generation of 3D+t cardiac meshes, we run Cardiac Mesh Flow and baseline methods to generate 1,000 mesh sequences. Between 1,000 generated cardiac mesh sequences and 500 real test mesh sequences, we compare the distributions of cardiac four-chamber volumes. In detail, we compute the volume $V_{\tau}^s$ for each of the five cardiac structures $s\in\{\mbox{LV, LVM, RV, LA, RA}\}$ and for each time frame $\tau=1,...,50$, resulting in a volume vector with 250 elements describing each heart. Similarly to CardiacFlow \citep{ma2025cardiacflow}, we then evaluate the generation fidelity and diversity using \textit{volume Fréchet inception distance} (vFID) \citep{heusel2017fid}, which is formulated as:
\begin{equation}\label{eq:vfid}
\mathrm{vFID}=\left\|\hat{\mu}-\mu_*\right\|^2+\mathrm{tr}\left(\hat{\Sigma}+\Sigma_*-2\left(\hat{\Sigma}\Sigma_*\right)^{\frac{1}{2}}\right),
\end{equation}
where $\hat{\mu},\mu_*\in\mathbb{R}^{250}$ and $\hat{\Sigma},\Sigma_*$ are the mean and covariance of the volume vectors of the generated and real cardiac mesh sequences, respectively. Such a vFID score measures the discrepancy of the four-chamber volume distributions between the synthetic and real cardiac mesh sequences. To rescale the vFID score to a valid range, all volumes are normalised by the end-diastolic (ED) volume of the heart, \emph{i.e.}, $\hat{V}_{\tau}^s=5\cdot V_{\tau}^s/(\sum_s V^s_1)$. 

\paragraph{Comparison results} The comparison results of Cardiac Mesh Flow with baseline generative models are reported in Table~\ref{tab:compare}. The flow matching-based models achieve consistently better vFID scores compared to VAE-based models \citep{gaggion2025hybridvnet,qiao2025meshheart,dou2025cardiosynth}. This is possibly because that VAE-based models are affected by the trade-off between the generation fidelity and diversity, tending to over-smooth the generated mesh sequences. Cardiac Mesh Flow substantially improves the vFID score and generates cardiac four-chamber meshes with realistic distribution of chamber volumes. Fig.~\ref{fig:uncond} visualises different samples of 3D+t cardiac meshes generated by Cardiac Mesh Flow with high quality and variability.

\paragraph{Computational cost} We report the inference time and GPU memory cost for all methods in Table~\ref{tab:compare}. MeshHeart \citep{qiao2025meshheart} is the most time-efficient method, as it captures cardiac motion by a temporal Transformer in the latent space and generates all frames of the 3D+t cardiac mesh sequence simultaneously. HybridVNet \citep{gaggion2025hybridvnet} and 4D CardioSynth \citep{dou2025cardiosynth} are more time-consuming since they use relatively large GCN models. Cardiac Mesh Flow has a marginally higher computational cost than one-step flow matching approach, due to the introducing of a multi-scale fusion network. Cardiac Mesh Flow can be further accelerated by generating a complete 3D+t cardiac mesh sequence in one batch, if GPU memory allows.

\begin{figure*}[h]
\centering
\includegraphics[width=0.985\linewidth]{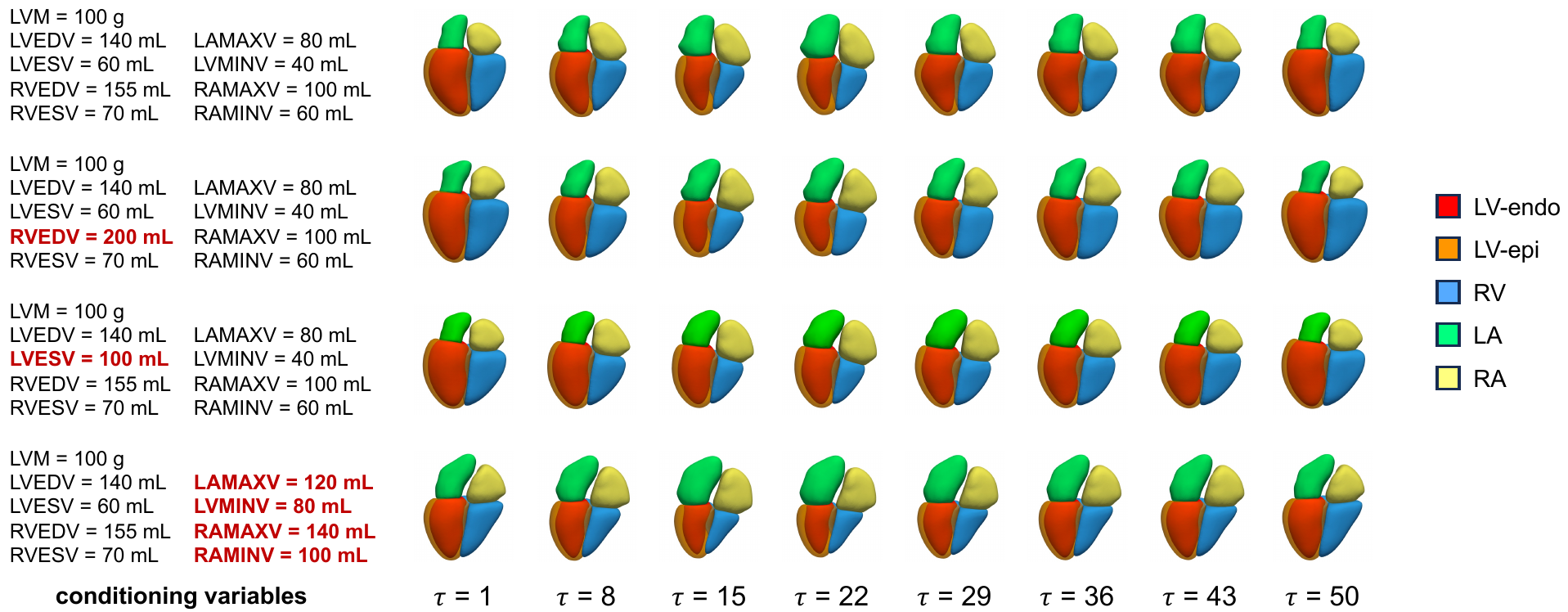}
\caption{Conditional generation of 3D+t cardiac four-chamber meshes using different conditioning variables. The different conditions are highlighted in red colour.}
\label{fig:cond}
\end{figure*}

\subsection{Conditional generation}
\paragraph{Evaluation metrics} For conditional generation, we use cardiac four-chamber phenotypes \citep{bai2020population} as input conditioning variables, including LV myocardial mass (LVM), ventricular end-diastolic and end-systolic volumes (LVEDV, LVESV, RVEDV, RVESV), and atrial maximum and minimum volumes (LAMAXV, LAMINV, RAMAXV, RVMINV). We examine if Cardiac Mesh Flow and baseline generative models can precisely control the phenotypes of the generated dynamic heart based on conditioning variables. We calculate the four-chamber phenotypes $c_*\in\mathbb{R}^9$ of all test data as the input conditioning variables to generate 3D+t mesh sequences. The conditional generation fidelity is measured by the root mean square deviation (RMSD) between the phenotypes $\hat{c}$ of the synthetic hearts and the corresponding input conditioning phenotypes $c_*$.

\paragraph{Comparison results} The RMSD errors for controllable conditional generation of all comparative approaches are reported in Table~\ref{tab:compare}. The flow-based approaches perform better than VAE-based methods \citep{gaggion2025hybridvnet,qiao2025meshheart,dou2025cardiosynth} conditioned on the four-chamber phenotypes. Cardiac Mesh Flow achieves the lowest RMSD between the generated and input conditioning phenotypes, providing accurate modulation of the synthetic 3D+t four-chamber meshes.

Furthermore, we provide qualitative visualisation in Fig.~\ref{fig:cond} for the conditional generation of 3D+t cardiac four-chamber meshes using different conditioning variables. With the increasing of the RVEDV, LVESV, and bi-atrial volumes, the generated cardiac meshes in Fig.~\ref{fig:cond} show larger RV size, weaker LV contraction, and larger atrial sizes respectively as expected. This demonstrates the advanced controllability and conditional generation ability of Cardiac Mesh Flow.

\begin{table*}[t]
\centering
\caption{The results of ablation experiments for Cardiac Mesh Flow regarding the inputs, temporal encoding methods, sampling strategies, and integration steps. The performance is evaluated using vFID score on unconditional generation. The periodic consistency is measured by the RMSD error between the mesh vertices of the first and last frames.}
\begin{tabular}{cccc|cc}
\toprule
Input & Temporal encoding & Sampling & Integration & vFID $\downarrow$ & $\mathrm{RMSD}_{\mathrm{period}}$ $\downarrow$ \\
\midrule
Embedding  & Periodic Gaussian ($\sigma=1$) & Uniform & 1-step & $17.01$ & $\mathbf{0.017}\pm\mathbf{0.002}$\\
Noise  & Periodic Gaussian ($\sigma=1$) & Beta & 1-step & $4.583$ & $0.239\pm0.048$\\
Embedding  & One-hot & Beta & 1-step & $0.471$ & $0.312\pm0.049$\\
Embedding  & Gaussian ($\sigma=1$) & Beta & 1-step & $0.502$ & $0.341\pm0.066$\\
Embedding  & Periodic Gaussian ($\sigma=1$) & Beta & 1-step & $\mathbf{0.439}$ & $0.283\pm0.050$\\
Embedding  & Periodic Gaussian ($\sigma=3$) & Beta & 1-step & $0.453$ & $0.181\pm0.036$\\
Embedding  & Periodic Gaussian ($\sigma=5$) & Beta & 1-step & $0.777$ & $0.136\pm0.030$\\
Embedding  & Periodic Gaussian ($\sigma=1$) & Beta & 2-step & $0.465$ & $0.282\pm0.051$ \\
Embedding  & Periodic Gaussian ($\sigma=1$) & Beta & 5-step & $0.490$ & $0.282\pm0.051$ \\
\bottomrule
\end{tabular}
\label{tab:ablation}
\end{table*}

\subsection{Ablation studies}
\paragraph{Evaluation metrics} We perform ablation studies on unconditional 3D+t cardiac four-chamber mesh generation to verify the effectiveness of each individual component of Cardiac Mesh Flow. The performance is assessed by the vFID score (\ref{eq:vfid}) as well as the periodic consistency of the generated 3D+t cardiac meshes \citep{ma2025cardiacflow}. For each mesh sequence, the periodic consistency is measured by the RMSD of all vertices between the meshes at the first ($\tau=1$) and last ($\tau=50$) time frame. The results are presented in Table~\ref{tab:ablation}.

\paragraph{One-step generation} We conduct ablation experiments on necessary elements of Cardiac Mesh Flow for one-step generation, including Beta sampling and input learnable embedding. Instead of using Beta sampling strategy to draw more time points near $t=0$, we sample the integration time $t$ from a uniform distribution $\mathcal{U}(0,1)$ during training like standard flow matching \citep{lipman2023fm,liu2023rectified}. As shown in Table~\ref{tab:ablation}, the uniform sampling leads to the lowest generation fidelity, while the highest periodic consistency is achieved as it generates static hearts and fails to capture the motion patterns.

Moreover, we replace the input learnable embedding with random Gaussian noise $\mathcal{N}(0,I)$. We then observe that Cardiac Mesh Flow fails to produce high fidelity cardiac mesh sequences within one step, as it is challenging to learn straight flow trajectories from Gaussian noise to multi-scale FFD fields. Additionally, we use more integration steps for inference, \emph{e.g.}, $K=2$ and $K=5$. However, the results in Table~\ref{tab:ablation} show that more integration steps do not necessarily improve the performance of Cardiac Mesh Flow, whereas one-step generation is sufficient to synthesise high quality cardiac meshes.

\paragraph{Temporal encoding} To validate the effectiveness of the periodic Gaussian kernel encoding $\mathcal{K}_{\sigma}$, we compare to the one-hot encoding and the regular Gaussian kernel encoding of the time frame $\tau$. Table~\ref{tab:ablation} demonstrates that the periodic Gaussian kernel encoding yields superior anatomical fidelity and periodic consistency than other temporal encoding strategies. The increasing of the standard deviation $\sigma$ results in further improvement of the periodic consistency. However, the generation quality will be affected by the temporal over-smoothness if the standard deviation is too large, \emph{e.g.}, $\sigma=5$.

\section{Conclusion}
In this work, we presented Cardiac Mesh Flow, a one-step flow matching approach for 3D+t cardiac four-chamber mesh generation. Cardiac Mesh Flow learned the distribution of multi-scale FFD fields, which warped a template mesh to generate realistic cardiac four-chamber anatomy at each time frame of a cardiac cycle. Our experiments showed that Cardiac Mesh Flow achieved high generation fidelity and diversity, outperforming SOTA cardiac mesh generative models based on VAE \citep{gaggion2025hybridvnet,qiao2025meshheart,dou2025cardiosynth}. In addition to unconditional generation, we demonstrated the controllability of Cardiac Mesh Flow using conditioning variables such as cardiac chamber volumes to control the generation. This envisages future generative modelling research for developing personalised cardiac digital twin models as well as patient-specific counterfactual analysis conditioned on physiological and clinical factors.

\section*{Declaration of Generative AI in the writing process}
During the preparation of this work, the author(s) used ChatGPT in order to improve language, grammar, clarity, and readability. After using this tool, the author(s) reviewed and edited the content as needed and take(s) full responsibility for the content of the publication.

\section*{Acknowledgments}
This research was conducted using the UK Biobank Resource under Application Number 18545. We thank all UK Biobank participants and staff. This work is supported by the EPSRC grants (EP/W01842X/1, EP/Z531297/1) and the BHF New Horizons Grant (NH/F/23/70013). P.M.M. acknowledges generous personal support from the Edmond J. Safra Foundation and Lily Safra, an NIHR Senior Investigator Award, Rosalind Franklin Institute, and the UK Dementia Research Institute, which is funded predominantly by the UKRI Medical Research Council. D.P.O. is supported by the Medical Research Council (MC\_UP\_1605/13), the NIHR Imperial College Biomedical Research Centre, and the British Heart Foundation (RG/F/24/110138, RE/24/130023, CH/F/24/90015).



\bibliographystyle{model2-names.bst}\biboptions{authoryear}
\bibliography{ref}



\end{document}